\documentclass[aps,prl,superscriptaddress,twocolumn,floatfix]{revtex4-1}
\usepackage{units}
\usepackage{amsmath}
\usepackage{amssymb}
\usepackage{graphicx}
\usepackage{bm}
\usepackage{multirow,color,relsize,ulem,microtype}

\newcommand{\be}{\begin{equation}}
\newcommand{\ee}{\end{equation}}
\newcommand{\e}{\varepsilon}

\newcommand{\re}[1]{\text{Re}[#1]}
\newcommand{\im}[1]{\text{Im}[#1]}

\newcommand{\cc}[1]{{\color{black}#1}}
\newcommand{\pt}{\cal PT}
\newcommand{\uv}[1]{\hat{\bm{#1}}}

\begin{document}

\title{Defect states emerging from a non-Hermitian flat band of photonic zero modes}

\author{Bingkun Qi}
\affiliation{\textls[-18]{Department of Engineering Science and Physics, College of Staten Island, CUNY, Staten Island, NY 10314, USA}}
\affiliation{The Graduate Center, CUNY, New York, NY 10016, USA}

\author{Lingxuan Zhang}
\affiliation{\textls[-18]{Department of Engineering Science and Physics, College of Staten Island, CUNY, Staten Island, NY 10314, USA}}
\affiliation{\textls[-40]{State Key Laboratory of Transient Optics and Photonics, Xi'an Institute of Optics and Precision Mechanics, Chinese Academy of Sciences, Xi'an 710119, China}}

\author{Li Ge}
\email{li.ge@csi.cuny.edu}
\affiliation{\textls[-18]{Department of Engineering Science and Physics, College of Staten Island, CUNY, Staten Island, NY 10314, USA}}
\affiliation{The Graduate Center, CUNY, New York, NY 10016, USA}

\date{\today}

\begin{abstract}
We show the existence of a flat band consisting of photonic zero modes in a gain and loss modulated lattice system, as a result of the underlying non-Hermitian particle-hole symmetry. This general finding explains the previous observation in parity-time symmetric systems where non-Hermitian particle-hole symmetry is hidden. We further discuss the defect states in these systems, whose emergence can be viewed as an unconventional alignment of a pseudo-spin under the influence of a complex-valued pseudo-magnetic field. These defect states also behave as a chain with two types of links, one rigid in a unit cell and one soft between unit cells, as the defect states become increasingly localized with the gain and loss strength.
\end{abstract}

\maketitle


Defect states are ubiquitous in periodic systems due to the existence of bandgaps. In the simple case of a point defect, if its energy falls deep into a bandgap, then it cannot couple efficiently to the rest of the system, where no propagating mode exists at its energy. As a result, a defect state localized at this point is formed, no matter whether the defect is in the bulk or at the edge of the system.
Take the simplest periodic system in one dimension (1D) for example, its unit cell contains one element of energy $\omega_0$ that couples to its nearest neighbors with strength $t>0$, where a single band extends from $[\omega_0-2t,\omega_0+2t]$ across the Brillouin zone (BZ). A defect state forms if the on-site energy of a single defect at the edge is detuned from $\omega_0$ by more than $t$, and it appears above (below) this band if the detuning is positive (negative).

A particular interesting case for defect states is in the presence of a flat band, where a small detuning is sufficient to create a defect state in general. A flat band is dispersionless inside the whole BZ, and systems that exhibit flat bands have attracted considerable interest in the past few years, including optical \cite{Manninen1,Manninen2} and photonic lattices \cite{Rechtsman,Vicencio,Mukherjee,Biondi}, graphene \cite{Kane,Guinea},  superconductors \cite{Simon,Kohler1,Kohler2,Imada}, fractional quantum Hall systems \cite{Tang,Neupert,Sarma} and exciton-polariton condensates \cite{Jacqmin,Baboux}. Due to the singular density of states at the flat band energy, several interesting localization phenomena and their scaling properties have been identified \cite{Chalker, Bodyfelt, Flach,Leykam,AdP17}.

In Refs. \cite{Flatband_PT,Chern,Molina}, parity-time ($\pt$) symmetric perturbations, i.e., those with a complex potential satisfying $V(x)=V^*(-x)$ 
\cite{Bender1,Bender2,Bender3,El-Ganainy_OL07,Moiseyev,Musslimani_prl08,Makris_prl08,Kottos,Guo,mostafazadeh,Longhi,CPALaser,conservation,conser2D,Robin,EP9,nonlinearPT,Ruter,Lin,Microwave,Feng_NM,Feng2,Walk,Hodaei,Yang,Chang}, were introduced to study their effects on an existing flat band in the underlying Hermitian system. In the meanwhile, it was known that the introduction of a $\cal PT$-symmetric potential can collapse two neighboring bands into a single one in terms of their real parts \cite{Makris_prl08}, which is flat in some cases \cite{PTSSH,Feng_SciRep}. The conditions that lead to this flatness in a non-Hermitian system were poorly understood, and in this work we point out that the mechanism that leads to these flatbands is actually due to another symmetry, i.e., non-Hermitian particle-hole (NHPH) symmetry \cite{Malzard,zeromodeLaser}.
We should mention that similar to the Hermitian case, a non-Hermitian flat band can also exist by engineering a Wannier function that is an eigenstate of the whole lattice [see Sec.~I in Supplementary Material (SM)].

With NHPH symmetry, the effective Hamiltonian anticommutes with an antilinear operator, and a particular simple way to achieve it employs a photonic lattice~\cite{zeromodeLaser}: starting with an underlying Hermitian system with chiral symmetry (also known as sublattice symmetry), which consists of identical elements on two sublattices coupled by nearest neighbor coupling (e.g., a square lattice, honeycomb lattice and so on), NHPH symmetry is automatically satisfied once spatial gain and loss modulation is applied.

The flat band resulted from NHPH symmetry consists of photonic zero modes, which share certain traits as their condensed matter counterparts (i.e., the Majorana zero modes \cite{Alicea,Sarma_QI,Beenakker}). However, these photonic zero modes are not necessarily localized in space, and we study the defect states emerging from these non-Hermitian flat bands by introducing a point defect. We employ the simplest 1D photonic lattice mentioned before but now with gain and loss modulation that doubles or quadruples the size of the unit cell. We show that a flat band is formed when the gain and loss strength $\gamma$ exceeds a critical value. Now by introducing a point defect at the edge of the system, a defect state appears and becomes increasingly localized as the non-Hermiticity of the system increases. This defect state behaves as a chain with two types of links, one rigid within a unit cell and one soft between unit cells. We find that the emergence of the defect state can be viewed as an unconventional alignment of a pseudo-spin under the influence of a complex-valued pseudo-magnetic field, and in some cases, the result of a $\pt$ transition. 
These results are first discussed using a tight-binding model and then verified by ab-initio vector simulations of Maxwell's equations in photonics waveguides.

\textit{Non-Hermitian Flat Band} --- The periodic system we consider is the simplest 1D lattice mentioned in the introduction, and we choose the identical on-site energy of the lattice sites to be the zero point of its energy levels. With the introduction of gain and loss modulation, the non-Hermitian system can be captured by the tight-binding model
\be
i\partial_t \psi_n = i\gamma_n \psi_n + t(\psi_{n-1} + \psi_{n+1}) \quad(n=1,2,\ldots).
\ee
Below we consider a periodic imaginary potential with $\gamma_n=\gamma_{n+m}$ where $m$ is an even integer. For an odd $m$ the system does not have two sublattices and hence NHPH symmetry does not hold.

\begin{figure}[t]
\centering
\includegraphics[width=\linewidth]{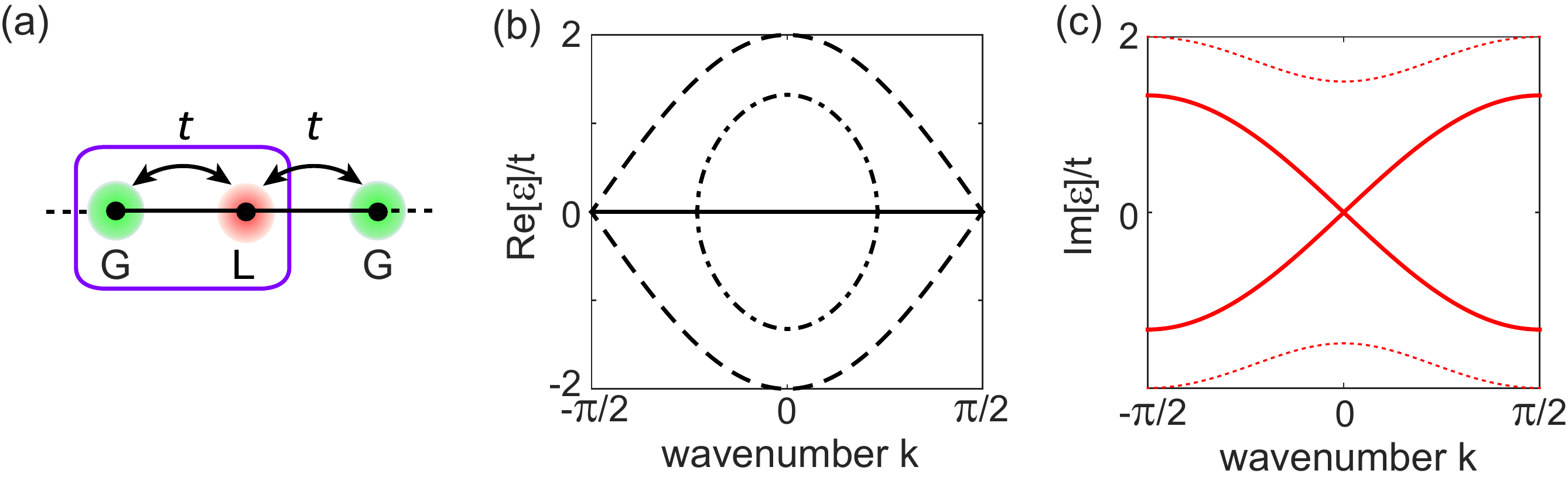}
\caption{(Color online) (a) Schematic of a gain and loss modulated lattice with period $m=2$. The box indicates the unit cell. (b) and (c) Real and imaginary parts of the bands in (a). The dashed lines in (b) mark the Hermitian bands when $\gamma=0$. The dash-dotted line show their partial collapse when $\gamma=1.5t$. The solid line shows the completed flat band when $\gamma\geq 2t$. The solid and dotted lines in (c) are for $\gamma=2t$ and $3t$, respectively. }\label{fig:band}
\end{figure}

When the period $m$ equals 2 [see Fig.~\ref{fig:band}(a)], the effective Hamiltonian can be written in the following form, by dropping an offset of the imaginary potential:
\be
H_2=
\begin{bmatrix}
i\gamma & t(1+e^{-2ik}) \\
t(1+e^{2ik}) & -i\gamma
\end{bmatrix}.\label{eq:H2}
\ee
$\gamma$ here is defined as $(\gamma_n-\gamma_{n+1})/2$, and we have set the distance between two neighboring lattice sites to be 1.
The dispersion relations of this system are then given by $\e_\pm(k) = \pm\sqrt{2t^2(1+\cos2k)-\gamma^2}$ in the BZ $k\in[-\pi/2,\pi/2)$. This effective Hamiltonian satisfies
\be
\{H_2,{\cal CT}\}=0,\; [H_2,{\cal PT}]=0,
\ee
i.e., it has both NHPH symmetry and $\cal PT$ symmetry (see Sec.~II in SM). Here $\cal T$ is the time-reversal operator in the form of the complex conjugation, and the chiral operator ${\cal C}=\sigma_z$ and parity operator ${\cal P}=\sigma_x$ are given by the Pauli matrices. The curly and square brackets denote anti-commutation and commutation relations as usual.

We note that $\cal PT$ symmetry dictates that the bands of the system satisfy $\e_i(k) = \e_j^*(k)$, where $i,j$ are band indices. In the case that $i,j$ are different, the two bands have the same $\re{\e}$ but different $\im{\e}$, which was a result of spontaneous $\cal PT$ symmetry breaking \cite{Bender2}. Nevertheless, $\cal PT$ symmetry does not ensure that their identical $\re{\e}$ needs to be flat in the BZ, and in Ref.~\cite{Makris_prl08} this merged band was indeed found to be curved.

NHPH symmetry, on the other hand, leads to a band structure satisfying $\e_i(k) = -\e_j^*(k)$ instead \cite{zeromodeLaser}. It clearly indicates that when $i=j$, a flat band at $\re{\e}=0$ can emerge with photonic zero modes. 
For the $m=2$ case above, this flat band starts to emerge from the boundary of the BZ as soon as $\gamma$ is nonzero, and it is formed completely when $\gamma>\gamma_c\equiv 2t$ [see Fig.~\ref{fig:band}(b)]. In \cc{Sec.~III} of SM we show another example where $m=4$ and the system lacks $\cal PT$ symmetry; the existence of a non-Hermitian flat band in this case corroborates the role of NHPH symmetry.

\begin{figure}[b]
\centering
\includegraphics[width=\linewidth]{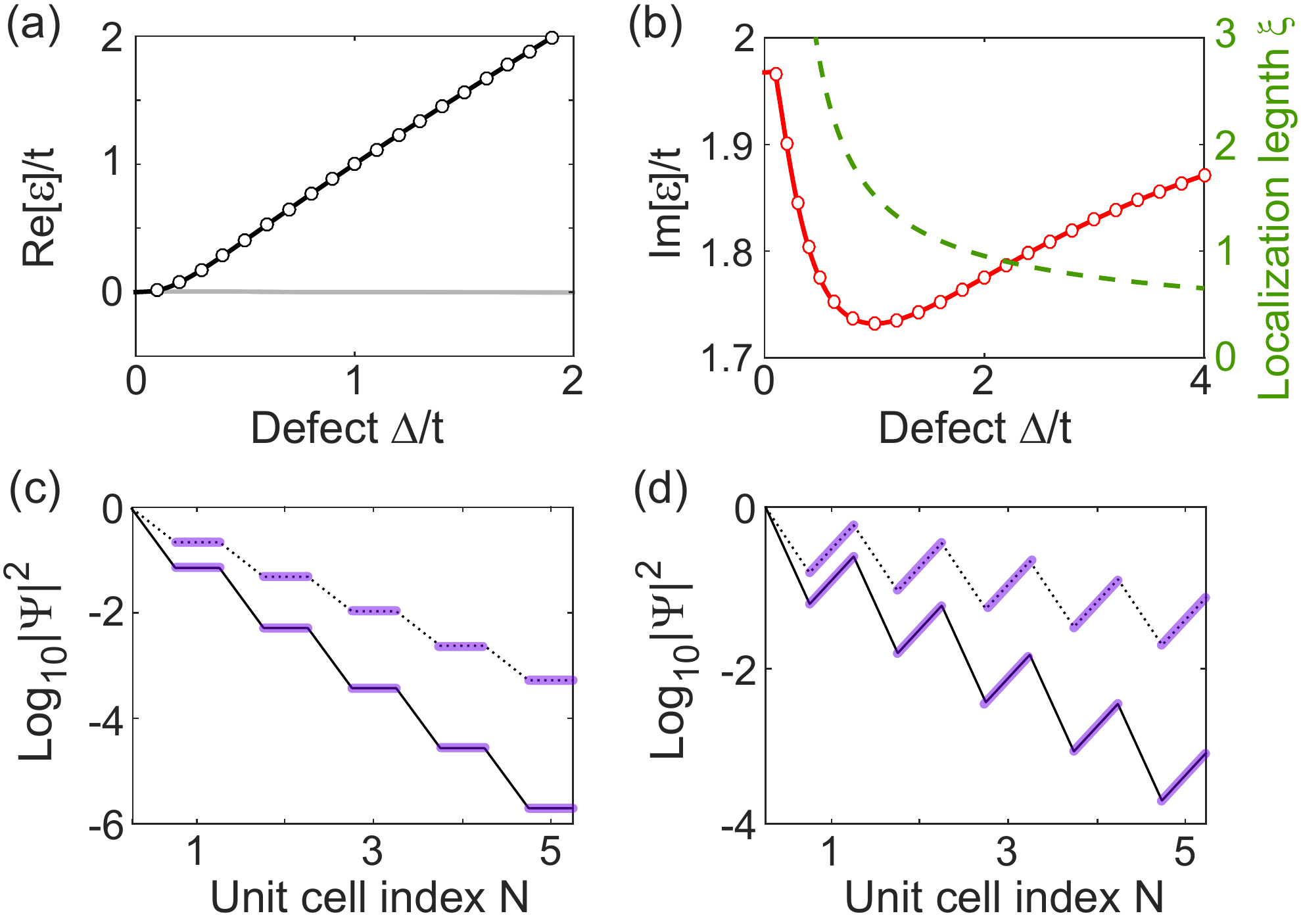}
\caption{(Color online) Emergence of a defect state from a non-Hermitian flat band as a function of the defect detuning $\Delta$, where the period of the gain and loss modulation is $m=2$. (a) and (b) Real and imaginary parts of the defect state energy as a function of the detuning $\Delta$. The solid lines and the dots show numerical results and the analytical expression (\ref{eq:defect}), respectively. $\gamma=2t$ is used. In (a) the grey lines show the almost unperturbed flat band energies of the bulk modes. In (b) the dashed line shows the localization length of the defect state. (c) Intensity profile of the defect state with $\Delta=t$. $\gamma=2t\,(1.3t)$ for the solid (dotted) line. Only the left 5 unit cells are shown (marked by the ``rigid links" that are parallel and $\gamma$-independent).
(d) Same as (c) but with $\Delta=t/2$. $\gamma=2t\,(1.1t)$ for the solid (dotted) line.
}\label{fig:defect}
\end{figure}

\textit{Defect States} --- Having shown that NHPH symmetry leads to a non-Hermitian flat band, next we probe the defect states emerging from it.
One example is shown in Fig.~\ref{fig:defect}(a) where a defect of detuning $\Delta$ is introduced to the left edge of the system (now of a finite length). We note that the defect state is formed at a small $\Delta$ as a result of the flat band, which is in contrast to the Hermitian case (e.g., the simplest 1D lattice) we have mentioned in the introduction.

One interesting feature of the defect state is its staggered intensity profile on the log scale [Figs.~\ref{fig:defect}(c) and (d)]: if we define the unit cells by counting from the $n=2$ site (i.e., avoiding the defect at the left edge), the intensity ratio $R$ within each unit cell is a constant for all unit cells. The same is true for the intensity ratio $R'$ between the gain (loss) sites in two neighboring unit cells.
Based on these observations, we derive an analytical expression for $\e_\Delta$ of the defect state in Sec.~IV of SM:
\be
\e_\Delta = \frac{(t^2+\Delta^2) \mp \sqrt{(t^2-\Delta^2-2i\gamma\Delta)^2+4t^2\Delta^2}}{2\Delta},\label{eq:defect}
\ee
where the ``$-(+)$" sign should be used for $\Delta<t\,(\Delta>t)$. This expression agrees nicely with the numerical data in Figs.~\ref{fig:defect}(a) and \ref{fig:defect}(b).

Furthermore, we find that the intra-cell intensity ratio $R$ mentioned above is simply given by
\be
R = \frac{\Delta^2}{t^2}
\ee
and \textit{independent} of the non-Hermitian parameter $\gamma$. In the meanwhile, the inter-cell intensity ratio $R'$ is given by
\be
R' = \frac{\Delta^4}{t^4}\left|\frac{\e_\Delta+i\gamma}{\e_\Delta-i\gamma}\right|^2,\label{eq:R'}
\ee
which does vary with $\gamma$. Therefore, the defect state behaves as a chain with two types links as we increase the non-Hermiticity of the system via $\gamma$, one rigid within a unit cell and one soft between unit cells.
This observation also indicates that the wave function of the defect state is exponentially localized on both sublattices [Figs.~\ref{fig:defect}(c) and (d)], with the \textit{same} localization length given by $\xi = 4/\ln R'$. At first glance this result may seem counterintuitive because
one would expect that the intensity of the wave will be amplified on the gain lattice sites and attenuated on the loss lattice sites, which will result in a varying inter-cell intensity ratio along the lattice and different localization lengths on the gain and loss sublattices. However, we remind the reader that here gain and loss do not describe wave propagation along the lattice. It is most obviously in a photonic lattice consisting of parallel waveguides, where the gain and loss characterizes wave propagation along the waveguides.
We also note that the localization length is not directly related to $\im{\e_\Delta}$. The latter is determined simultaneously by $R$ and $R'$, which lead to a non-monotonic $\Delta$-dependence of $\im{\e_\Delta}$ [see Fig.~\ref{fig:defect}(b)]; the localization length, on the other hand, reduces monotonically as $\Delta$ increases.

\begin{figure}[t]
\centering
\includegraphics[width=\linewidth]{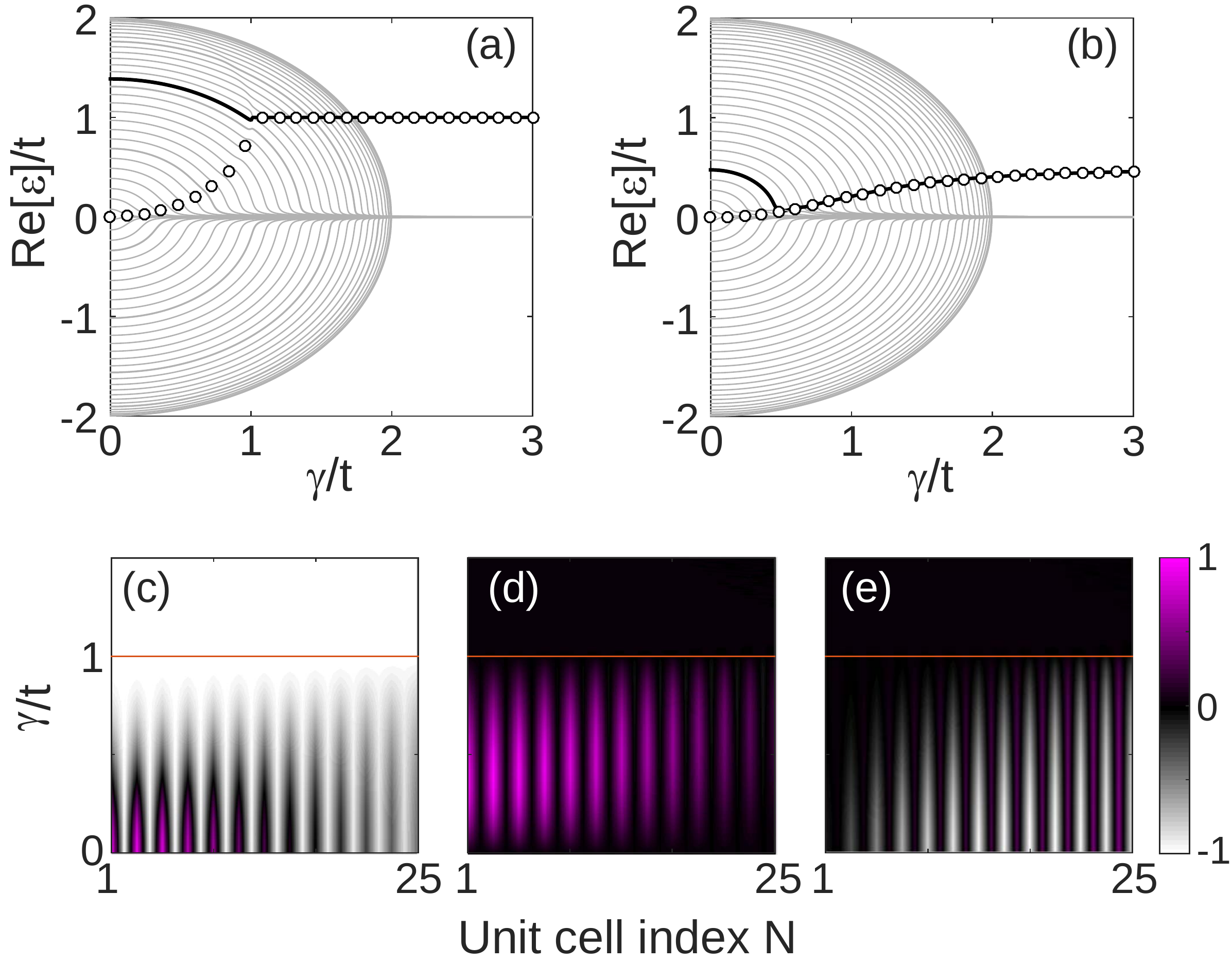}
\caption{(Color online) Emergence of a defect state from a non-Hermitian flat band as a function of the gain and loss strength $\gamma$ with period $m=2$. (a) and (b) Real part of all the modes in the system (solid lines) with $\Delta=t$ and $t/2$, respectively. The black line indicates the evolution of the defect mode, and the circles are the prediction of Eq.~(\ref{eq:defect}).
(c)--(e) False color plots of the pseudo-spin $\langle\sigma\rangle_{x,y,z}$ as a function of position and $\gamma$ in (a). Only the left 25 unit cells are shown.
}\label{fig:evol}
\end{figure}

Another interesting question about the defect state is how it evolves from the underlying Hermitian system as $\gamma$ increases and the flat band is formed. As Figs.~\ref{fig:evol}(a) and \ref{fig:evol}(b) show, the defect state originates from the middle of the Hermitian band, especially when $\Delta$ is small. By inspecting Eq.~(\ref{eq:defect}), we find that $|\Delta|=t$ is a special case, where a $\pt$ transition takes place at $\gamma=t$. We note that this is a different $\pt$ transition from those that take place on the real-$\e$ axis when the flat band is formed. We also note that Eq.~(\ref{eq:defect}) applies only when the defect state is localized and has a staggering intensity profile. Therefore, it is not surprising that its prediction in Fig.~\ref{fig:evol}(a) [and Fig.~\ref{fig:evol}(b)] deviates from the numerical result when $\gamma$ is small and the defect state is still in the bulk (see Sec.~V in SM). Nevertheless, the $\pt$-broken phase of $\e_\Delta$ in $\gamma>t$, characterized by its $\gamma$-independent real part, is faithfully manifested by the numerical data.

Now if we inspect the spatial profile of the defect state as it evolves with $\gamma$, we observe an unconventional alignment of a pseudo-spin under the influence of a complex-valued pseudo-magnetic field. To be more specific, we first rewrite the effective Hamiltonian (\ref{eq:H2}) using the Pauli matrices:
\be
H_2 = t(1+\cos ka)\,\sigma_x - t\sin ka\,\sigma_y + i\gamma\,\sigma_z \equiv -\bm{h}\cdot\bm{\sigma},
\ee
where $\bm{h}(\gamma)=[-t(1+\cos ka),\,t\sin ka, -i\gamma]$ is our complex-valued pseudo-magnetic field. We normalize the wave function $[\psi_L,\psi_G]^T$ in each unit cell when calculating $\langle\bm{\sigma}\rangle$, and the result is plotted in Figs.~\ref{fig:evol}(c)--(e) as a function of $\gamma$ when $\Delta=t$. It is clear that $\langle\bm{\sigma}\rangle$ displays a spatially dependent orientation when $\gamma<t$, but an aligned $\langle\bm{\sigma}\rangle$ is found across the whole lattice when $\gamma>t$. This value of $\langle\bm{\sigma}\rangle$ is given by $(-1,0,0)$ and can be viewed as the result an unconventional alignment of a pseudo-spin, since the direction of a complex $\bm{h}$ cannot be uniquely defined. The same alignment process takes place for other values of $\Delta$ as well. For example, $\langle\bm{\sigma}\rangle$ becomes $[-0.8,0,-0.6]$ when $\Delta=t/2$. We note that $\langle\sigma\rangle_y$ is always zero in the aligned state; it is in fact proportional to the optical flux between the gain and loss sites \cc{\cite{flux}} in a unit cell by definition [i.e., $i(\psi_G^*\psi_L-\psi_G\psi_L^*)$], which vanishes as one can show that $\psi_L/\psi_G=-\Delta/t$ is real (whose square gives $R$). Using this ratio we also derive $\langle\sigma\rangle_x=-2\Delta t/(\Delta^2+t^2)$, $\langle\sigma\rangle_z=(\Delta^2-t^2)/(\Delta^2+t^2)$, which agree nicely with their aforementioned numerical values (see also Sec.~VI in SM).

\begin{figure}[b]
\centering
\includegraphics[clip,width=\linewidth]{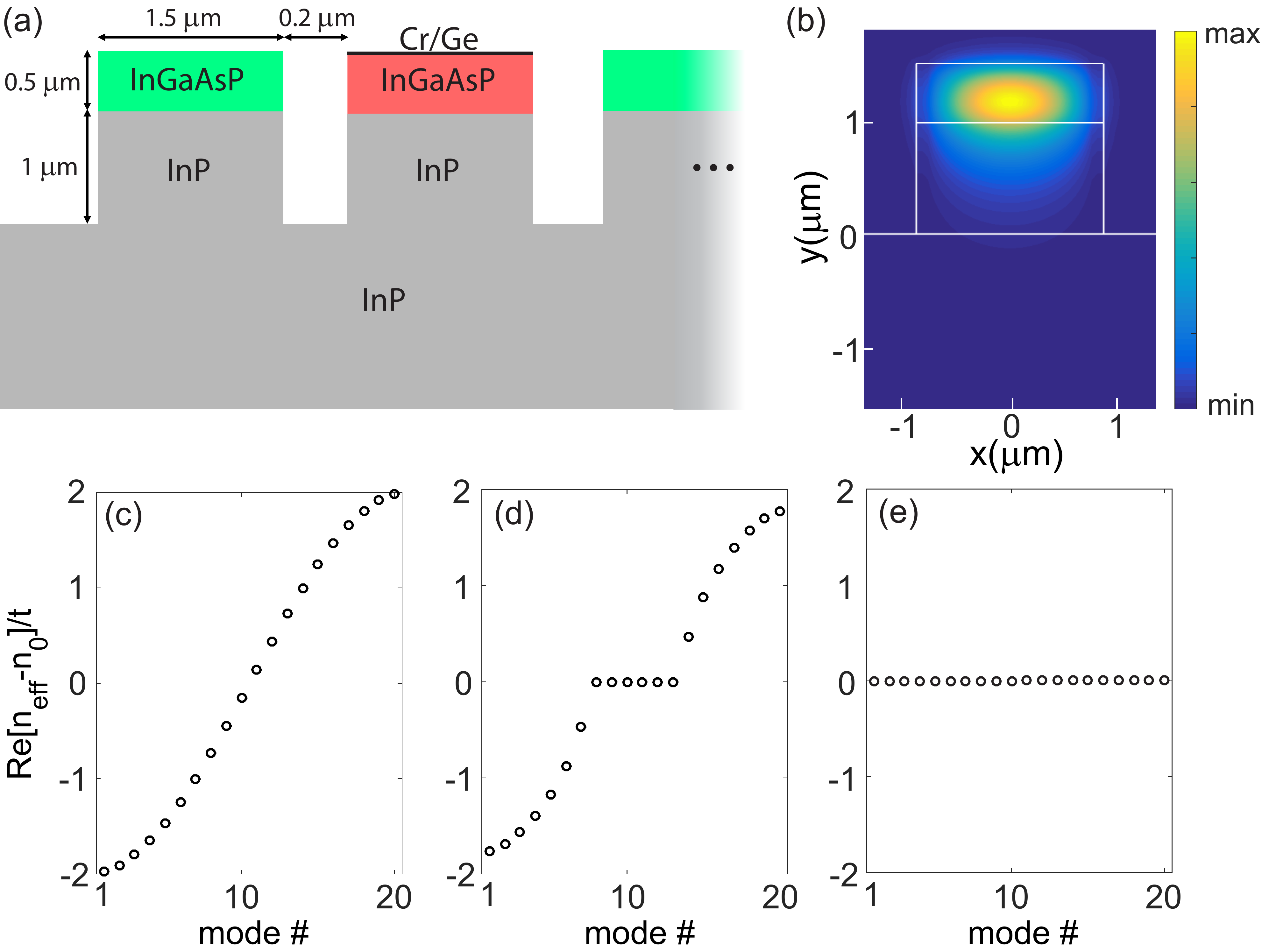}
\caption{(Color online) (a) Schematic of coupled photonic waveguides with alternate gain and loss. The refractive indices used are $3.17$ (InP), $3.44+in''$ (InGaAsP), and $3.44-in''$ (Cr/Ge $+$ InGaAsP). (b) $|E_x|$ component of the fundamental mode in a single waveguide when $n''=0$. (c)--(f) Real part of the band structure when $n''=0,t,2t$.}\label{fig:exp}
\end{figure}

\textit{Photonic realization} --- Next we present a realistic design using coupled photonic waveguides to demonstrate the practical feasibility of the predicted effects given above. Each waveguide has a square cross section, which is 1.5 $\mu$m wide and has a 500-nm-thick InGaAsP mutiple quantum wells on top of an InP substrate [see Fig.~\ref{fig:exp}(a)]. When optically pumped, the quantum wells supply the gain while the loss can be provided, for example, by a thin Cr/Ge double layer on top of the quantum wells that also blocks the pump. Similar structures have been used in a number of experimental demonstrations with fine controlled gain and loss ratios \cite{Miao,Wong}. The propagating mode along the waveguide direction can be denoted by $\vec{\Psi}(x,y,z)=\vec{E}(x,y)e^{-i\beta z}$, where $\vec{E}$ is the vector electric field. The propagation distance $z$ and propagation constant $\beta$ now play the roles of time and the eigenvalue $\e$ of the effective Hamiltonian, respectively.

\begin{figure}[t]
\centering
\includegraphics[clip,width=\linewidth]{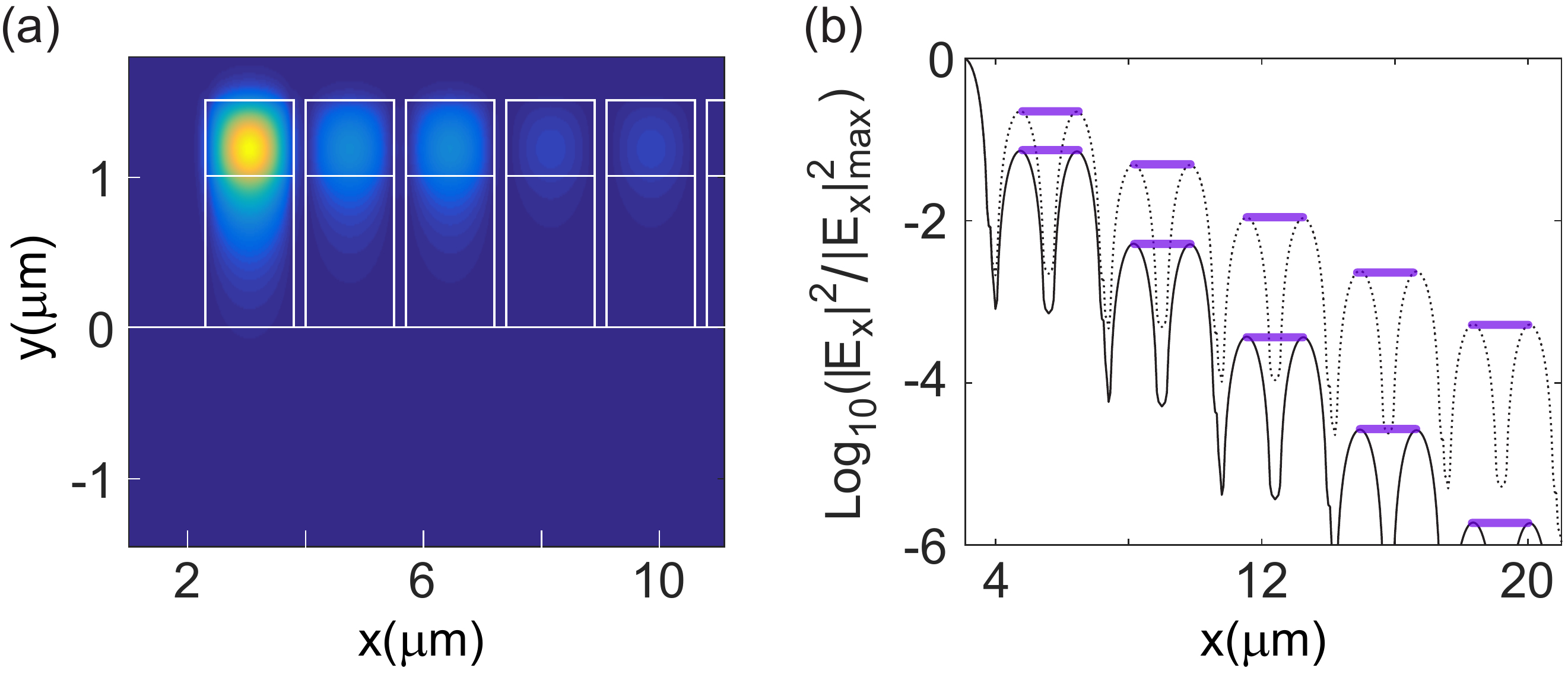}
\caption{(Color online) A defect state with a staircase profile. (a) Same as Fig.~\ref{fig:exp}(b) but with 20 coupled waveguides and $n''=2t$. The refractive index of the left waveguide is now increased by $\delta n = 7.43 \times10^{-5}$. (b) Its staggering profile at $n''=2t$ (solid line) and $1.3t$ (\cc{dotted} line). $|E_x|$ is taken at $y\approx1.2\,\mu$m where it is maximized. The horizontal ``rigid links" from Fig.~\ref{fig:defect}(c) are reproduced here and the agreement is excellent.}\label{fig:defect_FDTD}
\end{figure}

Below we introduce the effective index $n_\text{eff}=\beta\lambda/2\pi$ to characterize each propagating mode, with the wavelength chosen at $\lambda=1.55 \mu$m. By performing a finite-difference-time-domain simulation of Maxwell's equations using MEEP \cite{meep} and a perfectly matched layer as the global boundary condition, we find $n_\text{eff}=3.25\equiv n_0$ for the fundamental mode in a single waveguide [see Fig.~\ref{fig:exp}(b)]. With two coupled waveguides separated by 0.2 $\mu$m, we find that the two corresponding $n_\text{eff}$'s now differ by $1.17\times10^{-4}$, indicating a dimensionless coupling constant $t=5.83\times10^{-5}$. Now if we consider 20 waveguides, their individual fundamental modes couple to form a band with bandwidth $\Delta n_\text{eff}=2.31\times10^{-4}$, which agrees well with the tight-binding prediction ($4t$) mentioned in the introduction [see Fig.~\ref{fig:exp}(c)]. By increasing gain and loss incorporated as the imaginary part $n''$ of the top layer(s) that plays the role of the non-Hermitian parameter $\gamma$, we illustrate the forming of the non-Hermitian flat band in Figs.~\ref{fig:exp}(e) and \ref{fig:exp}(f) when $n''$ is increased to $2t$, again verifying the prediction of the tight-binding model. Furthermore, we introduce a ``point defect" similar to Fig.~\ref{fig:defect} by including an index detuning $\delta n = 7.43 \times10^{-5}$ in the gain layer of the left waveguide, which can be achieved, for example, by placing a layer of Ge on top of the waveguide \cite{Miao,Wong} (see also Sec.~VII in SM); it results in a change of the single-waveguide $n_\text{eff}$ by $t$, and we recover the staircase mode profile that displays an $n''$-independent ``rigid link" inside a unit cell and an $n''$-dependent ``soft link" between unit cells (see Fig.~\ref{fig:defect_FDTD}).

\textit{Conclusion and Discussion} --- In summary, we have shown that NHPH symmetry can lead to a flat band consisting of photonic zero modes, which explains the previous finding in $\pt$-symmetric systems where NHPH symmetry is hidden. Although we have only examined 1D lattices here, this mechanism also applies in higher dimensions (see Sec.~VIII in SM).
The defect states emerging from this flat band exhibit several interesting properties, such as possessing two types of links, one rigid within a unit cell and one soft between unit cells, as the defect states become increasingly localized with the non-Hermitian parameter. These behaviors, first \cc{predicted} using a tight-binding model, have been verified by full vector simulations of Maxwell's equations for the propagation modes in coupled photonic waveguides.

The emergence of these defect states can be viewed as an unconventional alignment of a pseudo-spin under the influence of a complex-valued pseudo-magnetic field, and in certain cases, the result of a $\pt$ transition. We note that for this pseudo-spin in our photonic lattice, spin-spin and spin-orbital interactions are absent and difficult to introduce, hence they are not considered here.

\appendix
\section{Supplementary Material}
\section{I. Another approach to generate a non-Hermitian flat band}

Besides the approach based on non-Hermitian particle-hole symmetry, there is another method to generate a non-Hermitian flat band, which follows the same recipe in a Hermitian system, i.e., engineering a Wannier function that is an eigenstate of the whole system. For example, we denote the Wannier function by $W_n(x-ja)$ in 1D, where $n$ is the band index, $a$ is the lattice constant, and $j$ is the unit cell index. The Bloch wave function with wave vector $k$ in the $n$th band can be written as
\be
\Psi_n(x;k) = \sum_j e^{ikaj}\,W_n(x-ja),
\ee
and it satisfies $H_0\Psi_n(x;k)=\omega_n(k)\Psi_n(x;k)$, where $H_0$ is the Hamiltonian of the entire system instead of the Bloch Hamiltonian $H(k)$ of the unit cell. Now if $H_0 W_n(x-ja) = \omega_w W_n(x-ja)$, i.e., if there exists an Wannier function that is an eigenstate of the whole system with eigenvalue $\omega_w$, then we immediately find $\omega_n(k)=\omega_w$ which is $k$-independent.

Typical this approach only applies to a limited number of frustrate lattices \cite{FB_Konotop,FB_Hami,FB_Yidong}, and it does not require the “phase transition” through an exceptional point or guarantee $\re{\e}=0$ (i.e., a flat band with photonic zero modes). Our approach based on the NHPH symmetry, on the other hand, is more general and reflects the true non-Hermitian nature of the gain and loss modulated lattices. A more detailed discussion will be presented elsewhere.

\section{II. NHPH and $\pt$ symmetries}

In the main text we have used the symmetry relations $\e_i(k)=-\e_j^*(k)$ and $\e_i(k)=\e_j^*(k)$ of the eigenvalue spectrum due to NHPH symmetry and $\pt$ symmetry respectively, where $i,j$ are band indices. Here we quickly review how these relations are derived.

NHPH symmetry is satisfied when the system Hamiltonian satisfies $\{H,{\cal CT}\}=0$, where $C$ is a linear operator and $\cal T$ is the time reversal operator. $\cal T$ takes the form of complex conjugation in our problem. If $\Psi_i$ is an eigenstate of $H$ with eigenvalue $\e_i$, i.e., $H\Psi_i=\e_i\Psi_i$, then $H({\cal CT}\Psi_i)=-{\cal CT}(H\Psi_i)=-{\cal CT}(\e_i\Psi_i)=-\e_i^*({\cal CT}\Psi_i)$. In the first step we have used the anti-commutation relation, and in the last step $\e_i$ acquires a complex conjugation when moved to the front of the $\cal CT$ operator ($\cal C$ does not act on $\e_i$, which is a single number). Therefore, ${\cal CT}\Psi_i$ is also an eigenstate of $H$ with eigenvalue $-\e_i^*$. We denote them by $\Psi_j$ and $\e_j$, which may or may not be the same as $\Psi_i$ and $\e_i$. When they are different, we have a pair of eigenvalues satisfying $\e_j=-\e_i^*$, which are in the broken NHPH phase, since $\Psi_j\equiv{\cal PT}\Psi_i\neq\Psi_i$; when they are the same, then we have $\e_i=-\e_i^*$, which means that $\re{\e_i}=0$; this is the NHPH-symmetric phase, because now we have ${\cal PT}\Psi_i=\Psi_i$.
The same conclusions hold when we replace the system Hamiltonian by the Bloch Hamiltonian, with which the eigenvalues and eigestates acquire a $k$-dependence.

Similarly, one finds that if $\e_i$ and $\Psi_i$ are one eigenvalue and eigenstate of $H$, then $\pt$ symmetry, given by $[H,\pt]=0$, leads to $H(\pt\Psi_i)=\pt(H\Psi_i)=-\pt(\e_i\Psi_i)=-\e_i^*(\pt\Psi_i)$. Therefore, $\Psi_j\equiv\pt\Psi_i$ is also an eigenstate of $H$ with eigenvalue $\e_j=\e_i^*$.

\section{III. With NHPH symmetry but without $\cal PT$ symmetry}

In the main text we have used the tight-binding model to illustrate the forming of a non-Hermitian flat band where each unit cell contains two gain and loss modulated lattice sites ($m=2$). In that case the system possesses both $\cal PT$ symmetry and NHPH symmetry.

\begin{figure}[b]
\centering
\includegraphics[width=\linewidth]{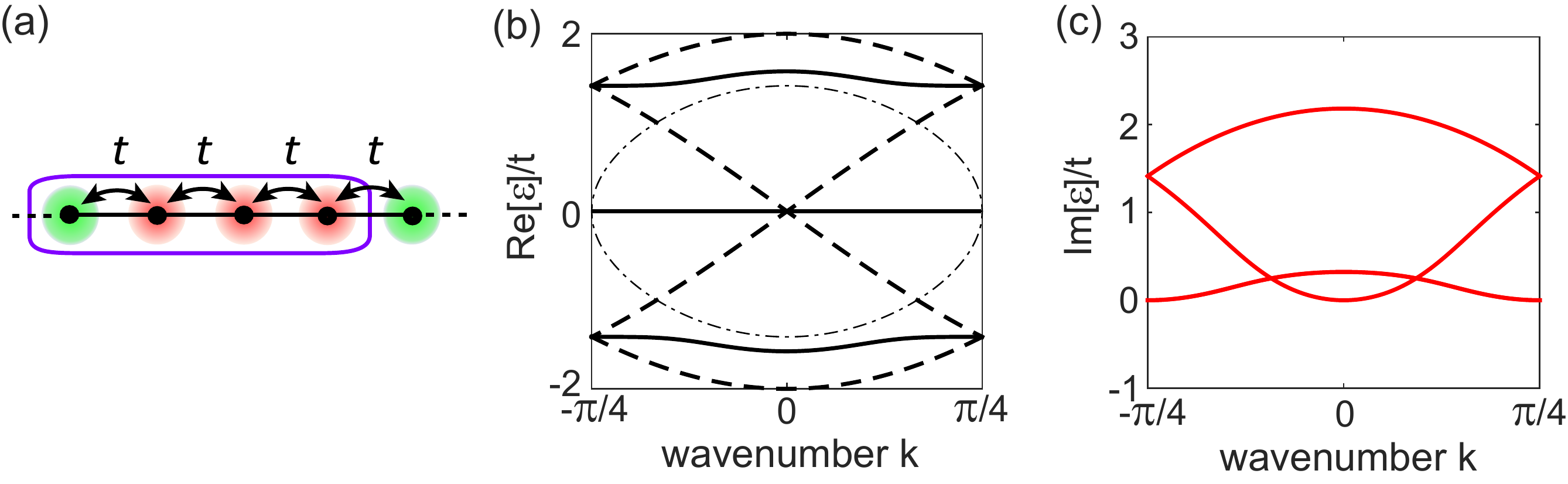}
\caption{\cc{(Color online) (a) Schematic of a gain and loss modulated lattice with period $m=4$. The box indicates a unit cell. (b) and (c) Real and imaginary parts of the bands in (a) when $\gamma_a=2\sqrt{2}t$ (solid lines). The dashed and dash-dotted lines in (b) show the Hermitian bands when $\gamma_a=0$ and the case $\gamma_a=\gamma_c=2\sqrt{2}t$, respectively.}}\label{fig:band4}
\end{figure}

Here we give another example where $m=4$ [see Fig.~\ref{fig:band4}(a)]. The effective Hamiltonian still satisfies NHPH symmetry but not $\cal PT$ symmetry in general. It can be written as
\be
H_4=
\begin{bmatrix}
i\gamma_a & t & 0 & te^{-4ik}  \\
t & i\gamma_b & t & 0\\
0 & t & i\gamma_c & t \\
te^{4ik} & 0 & t & 0
\end{bmatrix}
\ee
where $\gamma_{a,b,c} = \gamma_n-\gamma_{n+1,n+2,n+3}$. For simplicity, we consider just one nonzero $\gamma$ (e.g., $\gamma_a$). As $\gamma_a$ increases, the two central bands collapse into a flat band [Fig.~\ref{fig:band4}(b)], starting from the center of the BZ and completed when $\gamma_a\geq 2\sqrt{2}t$, again with photonic zero modes at $\re{\e}=0$. The upper and lower bands are then gapped from the flat band.
The lack of $\pt$ symmetry is obvious from the imaginary part of the bands shown in Fig.~\ref{fig:band4}(c), which would otherwise have a up-down symmetry. Due to the periodicity of the system, the same results hold if we use $\gamma_b$ or $\gamma_c$ as the nonzero non-Hermitian parameter.

Although we do not discuss the case of more than one nonzero $\gamma$ in detail, we mention that the two bandgaps next to the flat band shrink if we have a finite and positive $\gamma_c$, and they close completely when $\gamma_c=2\sqrt{2}t$ [see Fig.~\ref{fig:band4}(b)]. This tunability offers a flexible control of the non-Hermitian band structures, which the simple $\pt$-symmetric modulation of $m=2$ lacks.

\section{IV. Analytical results of the defect state}

In the main text we have derived an expression for the energy of the defect state
\be
\e_\Delta = \frac{(t^2+\Delta^2) \mp \sqrt{(t^2-\Delta^2-2i\gamma\Delta)^2+4t^2\Delta^2}}{2\Delta},\label{eq:defect}
\ee
as well as its intra-cell intensity ratio
\be
R = \frac{\Delta^2}{t^2}
\ee
and inter-cell intensity ratio
\be
R' = \frac{\Delta^4}{t^4}\left|\frac{\e_\Delta+i\gamma}{\e_\Delta-i\gamma}\right|^2.\label{eq:R'}
\ee
To derive these results, we denote the wave functions by $\Psi = [\psi^{(0)},\psi^{(1)}_L,\psi^{(1)}_G,\psi^{(2)}_L,\psi^{(2)}_G,\ldots]$ from the defect site on the left edge of the lattice to the loss and gain sites in the last unit cell. As mentioned in the main text, the wave function of the defect state has a staggering spatial profile, with which we assume a trial solution with the following property: $\alpha = \psi^{(1)}_G/\psi^{(0)} = \psi^{(n+1)}_G/\psi^{(n)}_G = \psi^{(n+1)}_L/\psi^{(n)}_L$. Plugging this trial solution into the tight-binding model given by Eq. (1) in the main text, i.e.,
\begin{align}
(i\gamma+\Delta)\psi^{(0)} + t\psi^{(1)}_L &= \e_\Delta\psi^{(0)}, \\
- i\gamma\psi^{(1)}_L + t(1+\alpha)\psi^{(0)}  &= \e_\Delta\psi^{(1)}_L, \\
i\gamma\psi^{(1)}_G + t(1+\alpha)\psi^{(1)}_L  &= \e_\Delta\psi^{(1)}_G,
\end{align}
we show below how to find $\alpha$, $\beta\equiv\psi^{(1)}_L/\psi^{(0)}$ and the eigenvalue $\e_\Delta$ of the defect state as a function of the defect strength $\Delta$.

We first rearrange the equations above into the following forms:
\begin{align}
t\beta &= (\e_\Delta-i\gamma-\Delta), \label{eq:tight1}\\
t(1+\alpha)  &= (\e_\Delta + i\gamma)\beta, \label{eq:tight2}\\
t(1+\alpha)\beta  &= (\e_\Delta - i\gamma)\alpha. \label{eq:tight3}
\end{align}
We then eliminate $\beta$ from these equations, by multiplying the two sides of Eqs.~(\ref{eq:tight1}, \ref{eq:tight2}) and Eqs.~(\ref{eq:tight2}, \ref{eq:tight3}), respectively:
\begin{align}
&t^2(1+\alpha) = (\e_\Delta+i\gamma)(\e_\Delta-i\gamma-\Delta),\label{eq:tight4}\\
&t^2(1+\alpha)^2 = (\e_\Delta^2+\gamma^2)\alpha,\label{eq:tight5}
\end{align}
Equation (\ref{eq:tight4}) is equivalent to
\be
\e_\Delta^2+\gamma^2 = (\e_\Delta+i\gamma)\Delta + t^2(1+\alpha),\label{eq:tight6}
\ee
which when substituted into the right hand side of Eq.~(\ref{eq:tight5}) gives
\be
\alpha = \frac{t^2}{(\e_\Delta+i\gamma)\Delta-t^2}.\label{eq:tight7}
\ee
Finally, by substituting $\alpha$ in Eq.~(\ref{eq:tight4}) by Eq.~(\ref{eq:tight7}), we derive a quadratic equation for $\e_\Delta$:
\be
\Delta(\e_\Delta+i\gamma) - t^2 = \frac{\Delta^2(\e_\Delta+i\gamma)}{\e_\Delta-i\gamma},\label{eq:tight8}
\ee
or equivalently,
\be
\Delta\, \e_\Delta^2 - (t^2+\Delta^2)\e_\Delta + \Delta\, \gamma^2 + i\gamma(t^2-\Delta^2) = 0,
\ee
which gives
\be
\e_\Delta = \frac{(t^2+\Delta^2) \mp \sqrt{(t^2-\Delta^2-2i\gamma\Delta)^2+4t^2\Delta^2}}{2\Delta}.
\ee
This is Eq.~(5) in the main text. In the case that $\Delta=t$, we find
\be
\e_\Delta = t \mp \sqrt{t^2-\gamma^2}, \label{eq:tight9}
\ee
and a $\pt$ transition happens at $\gamma=t$, which leads to the defect state shown in Fig.~4(a) in the main text as we have discussed there.

To derive the inter-cell intensity ratio $R$, we note
\be
R = \left|\frac{\beta^2}{\alpha^2}\right| = \left|\frac{\e_\Delta-i\gamma}{\e_\Delta+i\gamma}\frac{1}{\alpha}\right|,
\ee
where we have used the result of dividing Eq.~(\ref{eq:tight2}) from Eq.~(\ref{eq:tight3}) in the last step. By substituting $\alpha$ on the right hand side by Eq.~(\ref{eq:tight7}) and utilizing Eq.~(\ref{eq:tight8}), we find
\be
R = \frac{\Delta^2}{t^2},
\ee
which is Eq.~(6) in the main text.

To derive the intra-cell intensity ratio $R'$ given by Eq.~(7) in the main text, we note
\be
R' \equiv \frac{1}{|\alpha^2|}=\frac{\Delta^4}{t^4}\left|\frac{\e_\Delta+i\gamma}{\e_\Delta-i\gamma}\right|^2,
\ee
where we have used Eq.~(\ref{eq:tight7}) and again Eq.~(\ref{eq:tight8}).

\section{V. Small $\gamma$ behavior}

The results shown in Fig.~4 of the main text for a small $\gamma$ were obtained by solving numerically the tight-binding model given by Eq.~(1) in the main text, and from these numerical results we can trace the defect state back to its corresponding state in the bulk by reducing $\gamma$. For a small $\gamma$ where the defect state (as well as the flat band) has not yet formed, this corresponding bulk state is not strongly influenced by the point defect introduced to the left edge of the lattice. Therefore, to describe this bulk mode quantitatively, we may focus on how the increase of $\gamma$ induces the $\pt$ transition of this bulk mode (mode $+$) and its $\pt$-symmetric partner (mode $-$) in the absence of this point defect [As we mentioned in the main text and above, this is a different $\pt$ transition from that described by Eq.~(\ref{eq:tight9}) which takes place off the real axis.]

\begin{figure}[t]
\centering
\includegraphics[clip,width=.5\linewidth]{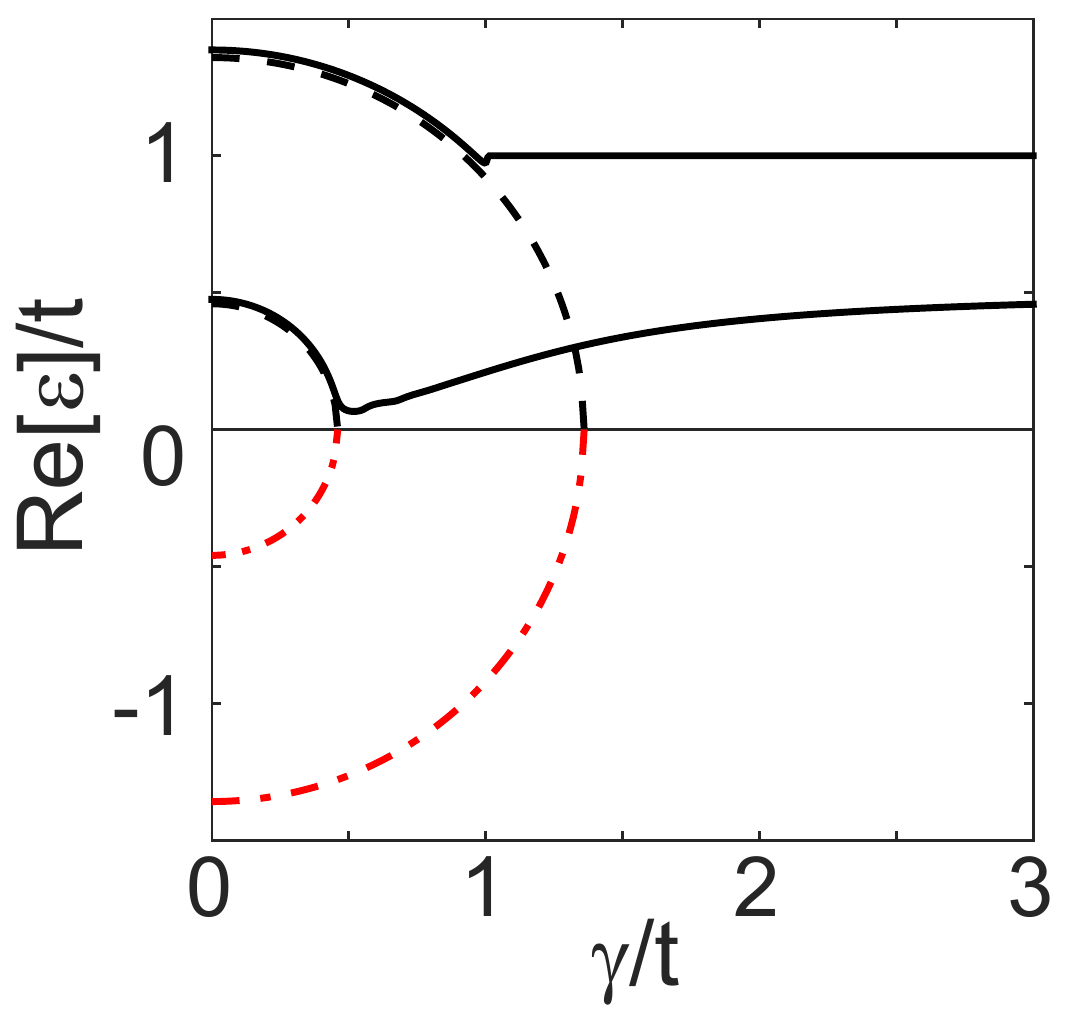}
\caption{Trajectories of the defect state (thick solid line) as a function of $\gamma$ with defect strength $\Delta/t=1$ (upper) and 0.5 (lower), adapted from Figs.~4(a) and 4(b) in the main text. Black dashed line and red dash-dotted line show the corresponding bulk state and its $\pt$-symmetric partner in the absence of the defect, up to the $\pt$ transition point $\gamma=|\e(\gamma=0)|$ on the real axis.
}\label{fig:semicircle}
\end{figure}

To capture this $\pt$ transition, we first note that the trajectories of modes $+$ and $-$ follow a semicircle in the $\gamma$-$\re{\e}$ plane for a positive $\gamma$ [Fig.~\ref{fig:semicircle}]. This is a very general behavior and does not depend on where modes $+$ and $-$ are in the bulk energy band. More specifically, we start with the Bloch Hamiltonian given by Eq.~(2) in the main text, whose band structure is given by
$\e_\pm (k;\gamma)= \pm\sqrt{\e_\pm^2(k;\gamma=0)-\gamma^2}$, where $\e_\pm^2(k;\gamma=0)=2t^2 (1+\cos⁡2ka)$ and $a$ is the lattice constant. It is then clear that for a given $k$ and a positive $\gamma$, the trajectories of $\e_\pm(k;\gamma)$ follow a semicircle in the $\gamma$-$\re{\e}$ plane:
\be
\e_\pm^2 (k;\gamma)+\gamma^2=2t^2 (1+\cos ⁡2ka).
\ee
We note that besides perturbing this general behavior, different values of $\gamma$ designate different bulk states to evolve into the defect state, as Fig.~\ref{fig:semicircle} below [as well as Figs.~4(a) and 4(b) in the main text] shows.

\section{VI. Physical meaning of the pseudo-spin}

In the main text we have shown that the emergence of the defect state can be viewed as an unconventional alignment of a pseudo-spin. Here we briefly discuss the physical meaning of the pseudo-spin. The pseudo-spin reflects the relative amplitude and phase of the wave functions on the loss and gain sites (denoted by $\psi_{L,G}^{(n)}$ and normalized by $|\psi_L^{(n)}|^2+|\psi_G^{(n)}|^2=1$) in a unit cell, and the alignment of the pseudo-spin means that the wave profiles in each unit cell become identical, including both the relative phase and amplitude of $\psi_{L,G}^{(n)}$. It does not, however, tell us the localization property of the defect state, which we have discussed using the inter-cell intensity ratio $R'$.

In the main text we have discussed the physical meaning of $\langle\sigma\rangle_y=i({\psi_G^{(n)}}^*\psi_L^{(n)} -c.c.)$, which is also the definition of the optical flux in a unit cell. We have mentioned that it vanishes in the defect state due to the real-valued ratio $\psi_L^{(n)}/\psi_G^{(n)}=-\Delta/t$. To be more clear, we now rewrite $\langle\sigma\rangle_y$ as
\be
\langle\sigma\rangle_y=2|\psi_L^{(n)}\psi_G^{(n)}|\sin⁡(\theta_G-\theta_L),
\ee
where $\theta_{L,G}$ are the phase angles of $\psi_{L,G}^{(n)}$. It is then clear that $\langle\sigma\rangle_y=0$ no matter whether $\Delta$ is positive or negative, with which the phase difference $\theta_G-\theta_L$ is either $\pi$ or 0.
Similarly, $\langle\sigma\rangle_x=2|\psi_L^{(n)}\psi_G^{(n)}|\cos⁡(\theta_G-\theta_L)$ also reflect the relative phase between $\psi_{L,G}^{(n)}$, and hence we have skipped its discussion in the main text.

Now
\be
\langle\sigma\rangle_z=\frac{\left|\psi_L^{(n)}\right|^2 - \left|\psi_G^{(n)}\right|^2}{\left|\psi_L^{(n)}\right|^2 + \left|\psi_G^{(n)}\right|^2} = \frac{R-1}{R+1},
\ee
where we have explicitly inserted the normalization condition $\left|\psi_L^{(n)}\right|^2 + \left|\psi_G^{(n)}\right|^2=1$ to show that $\langle\sigma\rangle_z$ reflects the relative amplitude of $\psi_{L,G}^{(n)}$, or equivalently, the intra-cell intensity ratio $R$.
Using the ratio $\psi_L^{(n)}/\psi_G^{(n)}=-\Delta/t$ again, we have also derived the analytical expressions of

\noindent $\langle\sigma\rangle_{x,z}$ given in the main text, i.e.,
\vspace{-4mm}
\be
\langle\sigma\rangle_x=-\frac{2\Delta t}{\Delta^2+t^2},\quad \langle\sigma\rangle_z=\frac{\Delta^2-t^2}{\Delta^2+t^2}.
\ee
Note that since $\langle\sigma\rangle_y=0$, we find $\langle\sigma\rangle_x^2+\langle\sigma\rangle_z^2=1$ in the defect state.

\section{VII. Point defect by a mirror plane}

\begin{figure}[t]
\centering
\includegraphics[clip,width=\linewidth]{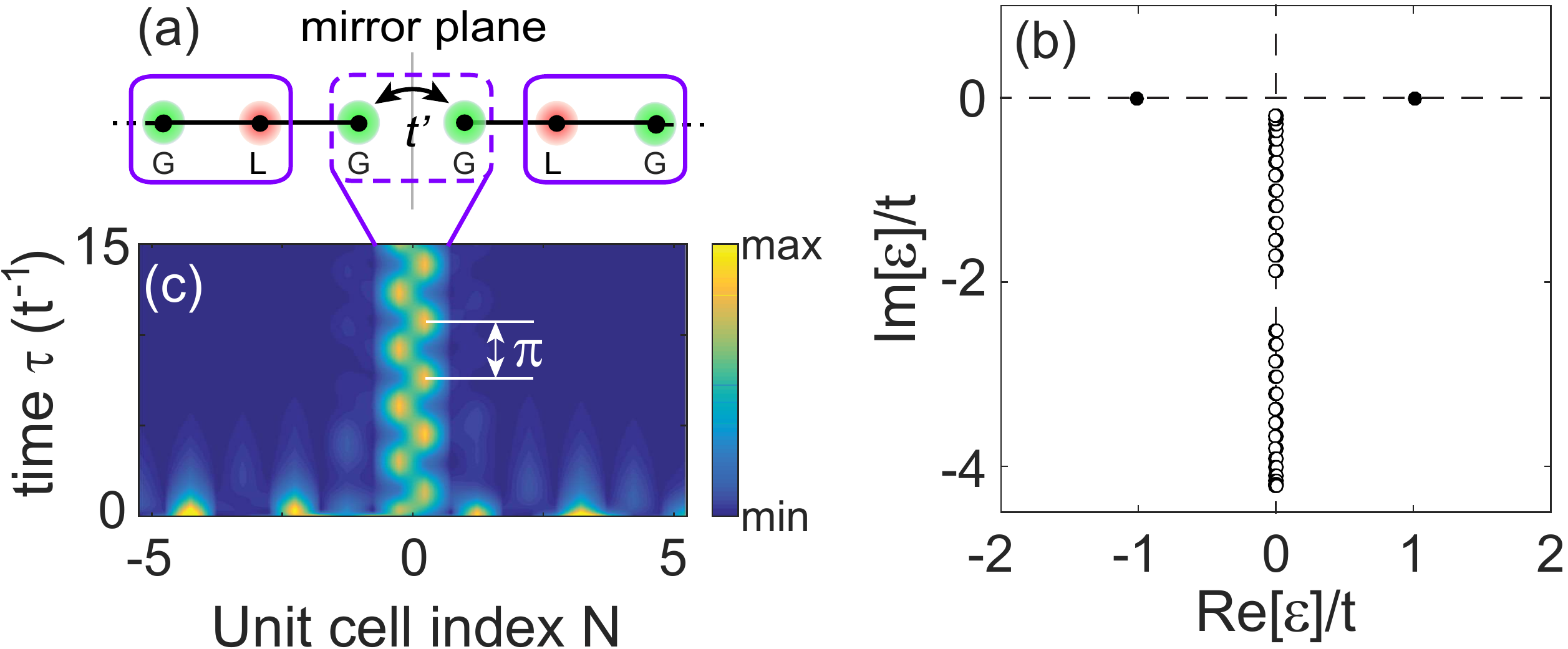}
\caption{\cc{(Color online) (a) Schematic of a symmetric setup with a non-Hermitian flat band. (b) Its spectrum at the laser threshold when $t'=t$, $\kappa=2.2t$ and $\gamma=2.02t$. The gain at the two central sites are 20\% stronger than the rest. The filled and open dots show the defect states and the bulk states in the perturbed flat band, respectively. (c) Temporal evolution of the laser at its threshold with an initial random noise.}}\label{fig:mirror}
\end{figure}

To introduce a point defect in a 1D lattice, one convenient way is to create a mirror plane as we show in Fig.~\ref{fig:mirror}(a) for the $m=2$ case: the even-parity modes of the system have an effective detuning of $t'$ at the two lattice sites right next to the mirror plane, where $t'$ is the coupling coefficient between the two halves of the system. Likewise, the odd-parity modes acquire an effective detuning of $-t'$. As a result, the defect states we have discussed now appear in pairs, one above the flat band and one below. They are NHPH-symmetric partners satisfying $\e_{+}=-\e_{-}^*\equiv\e_\Delta$, and they have an identical intensity profile and hence the same localization length. The latter can be controlled either by the non-Hermitian parameter $\gamma$ (i.e., how strong the gain and/or loss modulation is) or the effective detuning $\pm t'$ via the distance between the two halves.

To observe these defect states, one approach is to bring them to their lasing threshold ($\im{\e}=0$). In this setup we need to consider the intrinsic optical loss and absorption on each lattice site, and we take them to be uniform, represented by $-i\kappa$ on the diagonal of the effective Hamiltonian. There is one issue here though: because the defect states feel a stronger loss than the bulk states that reside mostly on the gain sublattice, the latter will reach their lasing thresholds before the defect states. Take the case shown in Fig.~2(b) of the main text for example, there are bulk states with $\im{\e}\approx\gamma>\im{\e_\Delta}$. To overcome this issue, we introduce a non-Hermitian defect, e.g., by making $\Delta$ complex and having a stronger gain. One example is shown in Fig.~\ref{fig:mirror}(b), and the pair of defect states in this symmetric setup indeed reach their lasing threshold before the bulk states. Assuming an inhomegeneous gain medium that supports both defect states, we can observe the blinking of the laser as a result of the beating between these two defect states, with a period given by $\tau=\pi/\re{\e_\Delta}$ [see Fig.~\ref{fig:mirror}(c)].

\section{VIII. Higher dimensions}

For the scenario discussed in our manuscript, i.e., two dispersive bands collapse with increasing non-Hermitian perturbation and form a band with the same $\re{\e}$, NHPH symmetry is a necessary and sufficient condition for the existence of a flat band also in 2D and 3D. Below we first exemplify the existence of such a flat band due to NHPH symmetry in 2D using a square lattice and 3D using a cubic lattice.

\begin{figure}[t]
\centering
\includegraphics[clip,width=\linewidth]{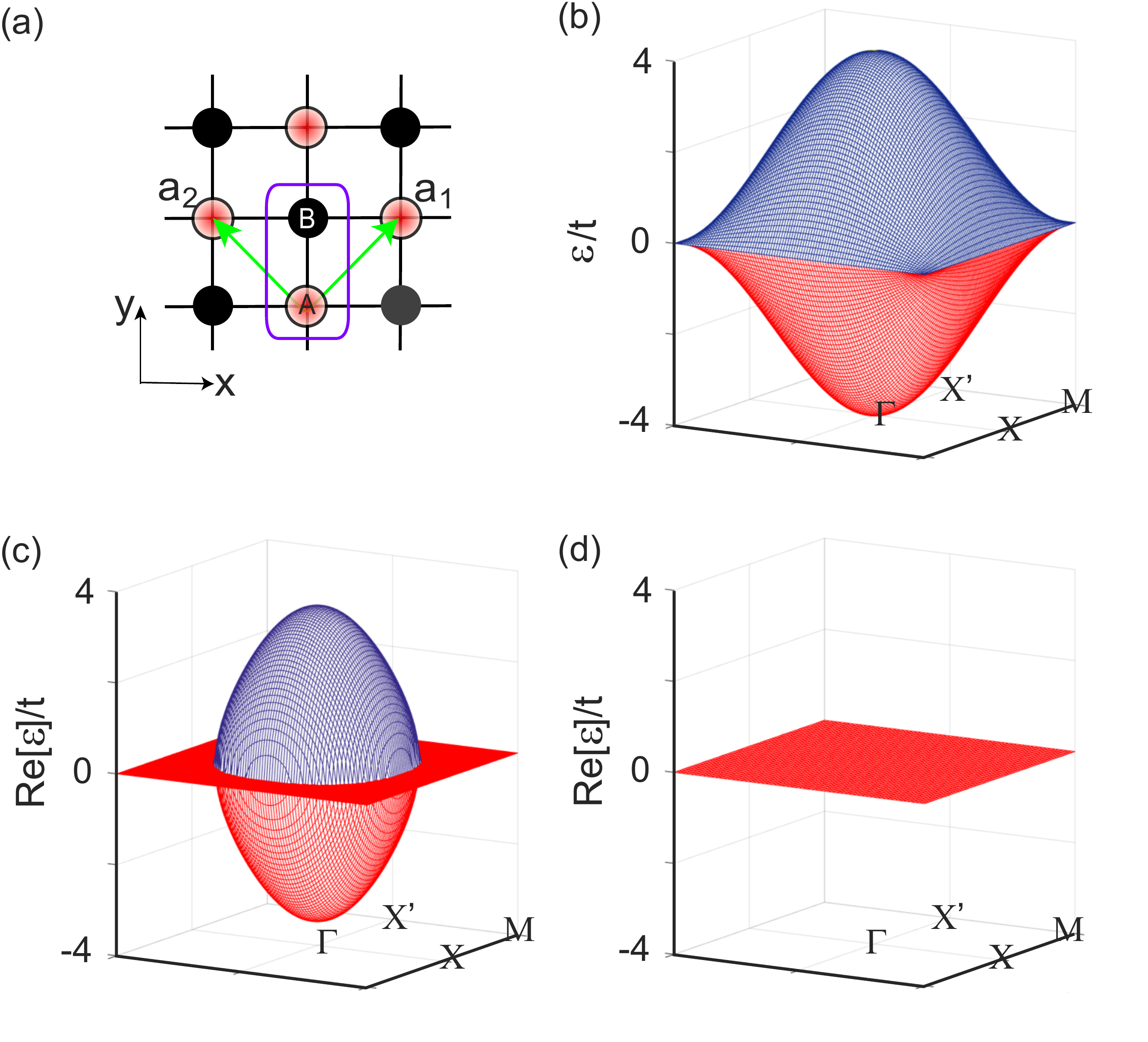}
\caption{Forming a non-Hermitian flat band in a square lattice. (a) Loss is introduced to the A sublattice. The green arrows show the primitive vectors $\bm{a}_{1,2}$. (b) Band structure in the Hermitian limit. The symmetry points $\Gamma,X,M,X'$ are marked in the Brillouin zone. (c)-(d) Real part of the band structure at $\gamma/t=-4$ and $-8$, respectively.}\label{fig:m2_2D}
\end{figure}

The square lattice we consider has two sublattices [imagine those of a checkerboard; marked by A and B in Fig.~\ref{fig:m2_2D}(a)] and they are coupled by nearest neighbor coupling $t$. We introduce the same amount of loss $\gamma$ to the A sublattice, and the unit cell consists of one A and one B lattice site. Its structure can be described by the primitive vectors $\bm{a}_{1,2}=a(\pm\uv{x}+\uv{y})$, where $a$ is the lattice constant of the underlying Hermitian lattice. The primitive vectors of the reciprocal lattice are given by $\bm{b}_{1,2}=\pi(\pm\uv{x}+\uv{y})/a$, satisfying $\bm{b}_i\cdot\bm{a}_j=2\pi\delta_{ij}\,(i,j=1,2)$.

In the Hermitian limit the two bands of the square lattice touch on the edges of the Brillouin zone [Fig.~\ref{fig:m2_2D}(b)], where the effective coupling between the two sublattices vanishes. As $|\gamma|$ increases, the flat band starts to form near the band edges [Fig.~\ref{fig:m2_2D}(c)], similar to the 1D case we have discussed in Fig.~1 of the main text. The flat band is completed when $|\gamma|=8t$ [Fig.~\ref{fig:m2_2D}(d)].

\begin{figure}[t]
\centering
\includegraphics[clip,width=\linewidth]{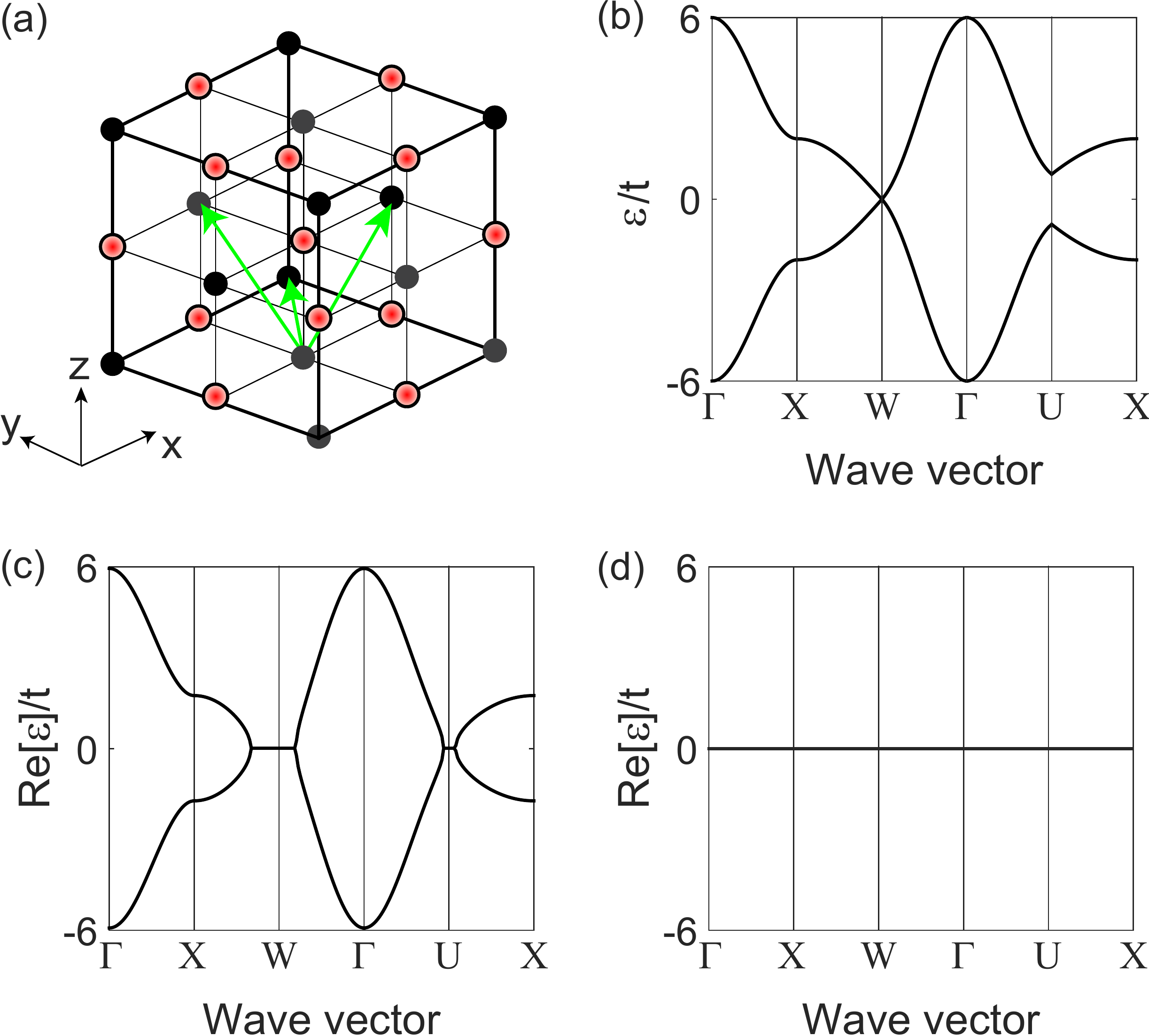}
\caption{Forming a non-Hermitian flat band in a cubic lattice. (a) The loss is introduced uniformly to one of the faced-centered cubic sublattice. The green arrows show the primitive vectors $\bm{a}_{1,2,3}$. (b) Band structure in the Hermitian limit. (c)-(d) Real part of the band structure at $\gamma/t=-2$ and $-12$, respectively.}\label{fig:m2_3D}
\end{figure}

The 3D example we consider is formed by stacking layers of the 2D lattice above, with each layer shifted by one lattice constant in both the $x$ and $y$ directions [Fig.~\ref{fig:m2_3D}(a)]. The resulting structure is that of the sodium chloride crystal, now with the ``sodium ions" having identical loss $\gamma$. Using the primitive vectors $\bm{a}_1=a(\uv{z}+\uv{x})$, $\bm{a}_2=a(\uv{x}+\uv{y})$, $\bm{a}_3=a(\uv{y}+\uv{z})$
of the lattice and $\bm{b}_1=\pi(\uv{z}+\uv{x}-\uv{y})/a$, $\bm{b}_2=\pi(\uv{x}+\uv{y}-\uv{z})/a$, $\bm{b}_3=\pi(\uv{y}+\uv{z}-\uv{x})/a$ of the reciprocal lattice, 
we calculate its band structure in the first Brillouin zone along the path $\Gamma$-$X$-$W$-$\Gamma$-$U$-$X$ \cite{bzpath}.
Figure ~\ref{fig:m2_3D}(b) shows its Hermitian limit, where the two bands are degenerate at the $W$ point. As $|\gamma|$ increases, the two bands start collapsing in the vicinity of $W$ and $U$ [Fig.~\ref{fig:m2_3D}(c)] and the flat band is completed when $|\gamma|=12t$.

\begin{figure}[b]
\centering
\includegraphics[clip,width=\linewidth]{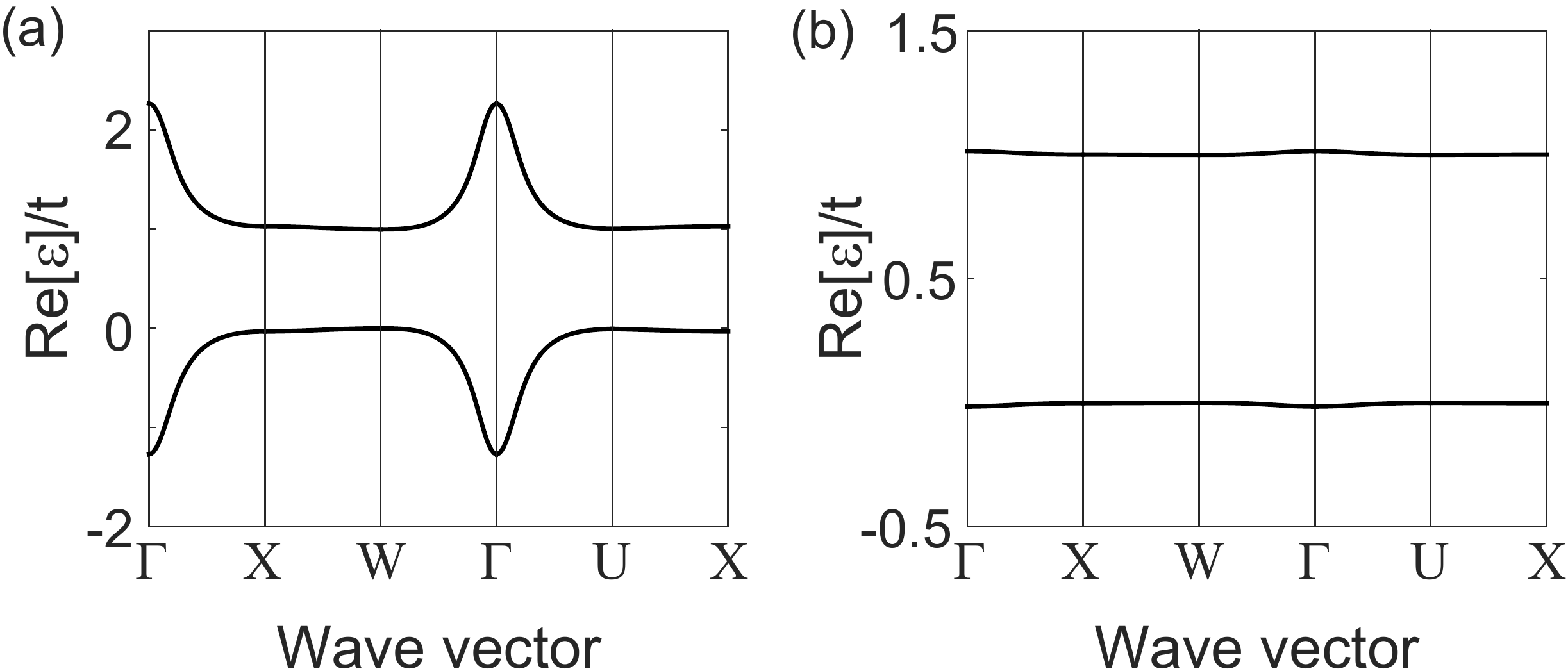}
\caption{Disappearance of a non-Hermitian flat band due to the lift of NHPH symmetry. (a) Same as Fig.~\ref{fig:m2_3D}(d) but with detuning $\Delta=t$. (b) Same as (a) but with $\gamma=-50t$.}\label{fig:m2_3D_2}
\end{figure}

Having shown that NHPH symmetry leads to a non-Hermitian flat band in 2D and 3D, below we lift the NHPH symmetry in the 3D example above and show the disappearance of the flat band. For this purpose, we introduce a detuning $\Delta$ on the ``sodium ions" where the loss has been introduced. As Fig.~\ref{fig:m2_3D_2}(a) shows, the flat band in Fig.~\ref{fig:m2_3D}(d) is again separated into two dispersive bands. We note that in the limit $|\Delta|\rightarrow\infty$ or $|\gamma|\rightarrow\infty$, the ``sodium ions" and ``chloride ions" are decoupled from each other. Since in the tight-binding model there is also no coupling between two ``sodium ions" (and two ``chloride ions"), two artificial flat bands, one residing on a sublattice, are approached in these limits [Fig.~\ref{fig:m2_3D_2}(b)]; they are artificial because for each periodic sublattice, every lattice point is isolated from the rest and there is no transport of any kind.

\begin{figure}[b]
\centering
\includegraphics[clip,width=\linewidth]{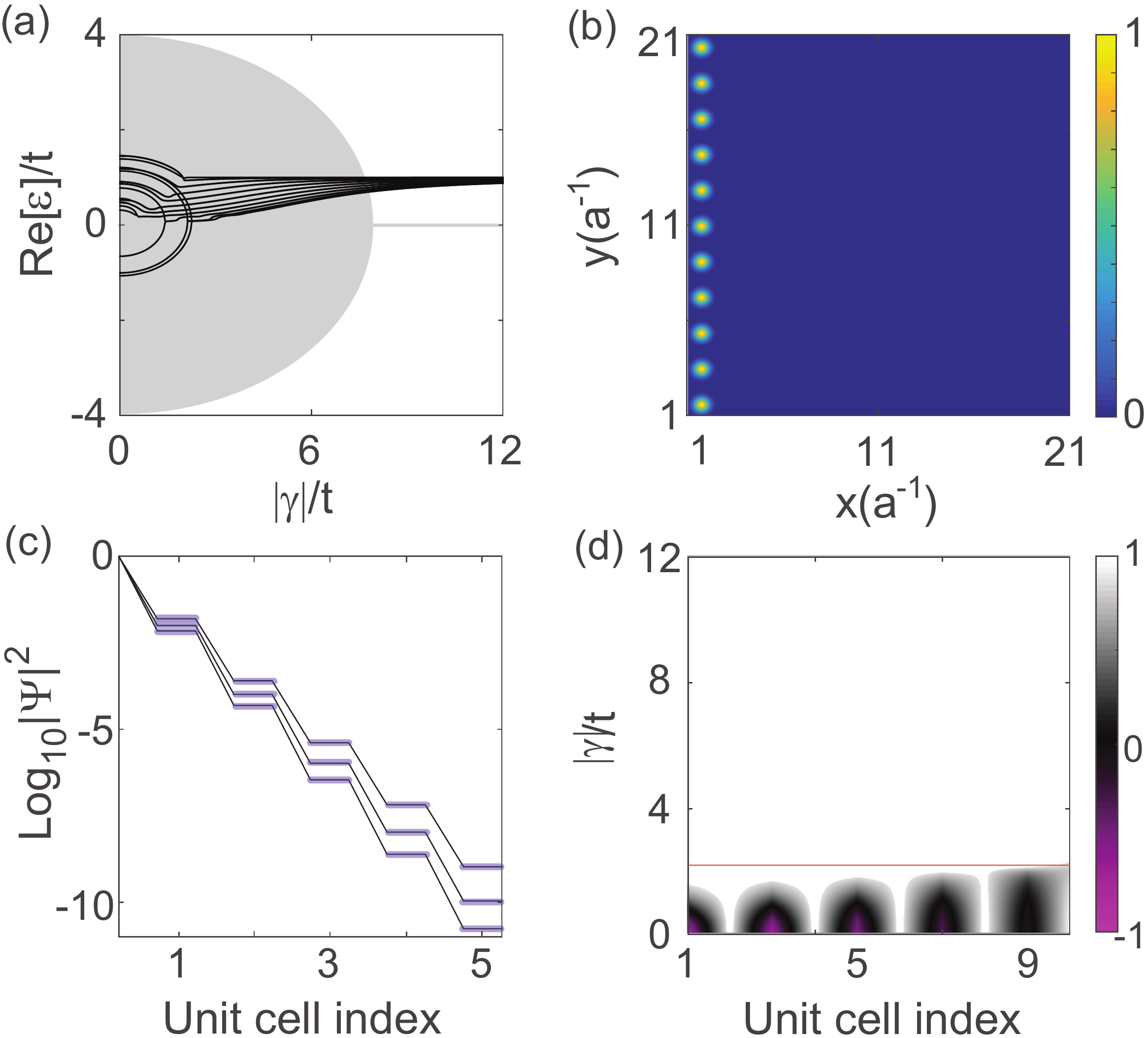}
\caption{Emergence of defect states in a square lattice. (a) Real part of the energy eigenvalues as a function of the loss strength $\gamma$. The detuning on the left edge of the lattice is $\Delta=t$. The grey region shows where the non-defect states exist and flat band they form. (b) False color plot of $|\Psi|^2$ in the defect state with the highest $\re{\e}$ at $\gamma=-8t$. (c) Staggering profile and two types of ``links" along a loss row. $\gamma/t=-8,-10,-12$ from top to bottom. The rigid links are marked by the purple sections. (d) $\langle\sigma\rangle_x$ along a loss row as a function of $\gamma$.}\label{fig:m2_2D_defect}
\end{figure}

In the last example of our higher dimensional discussions, we show the emergence of defect states in the 2D square lattice by introducing an edge defect. More specifically, we introduce the same detuning $\Delta$ on the left edge of the system. We note that the system is non-separable in the $x$-  and $y$-directions due to the checkerboard loss configuration, and hence the result we present below is not a trivial extension of the 1D case we have discussed in the main text. As Fig.~\ref{fig:m2_2D_defect}(a) shows, now $N_y$ (instead of 1) defect states emerge from the bulk of the system and are separated from the flat band as the loss strength $\gamma$ increases. Here $N_y=21$ is the number of grid points in the $y$-direction. Among these $N_y$ defect states, one bears a particular resemblance to the 1D defect state we have discussed in the main text: it has the same staggering wave function in the rows that have loss on the left edge (``loss rows"), and it is zero in the other rows [Fig.~\ref{fig:m2_2D_defect}(b)]. The two types of ``links" along each loss row are illustrated in Fig.~\ref{fig:m2_2D_defect}(c) with different values of $\gamma$, and the alignment of the pseudo-spins is exemplified by their $x$-component in Fig.~\ref{fig:m2_2D_defect}(d). The threshold value of the alignment is at $|\gamma|/t=2$ and marked by the red horizontal line; if we had introduced gain of the same strength on the other sublattice, this threshold value becomes 1 as is the case in Fig. 4(c) of the main text.

\end{document}